\documentclass[aps,prl,amsmath,superscriptaddress,twocolumn,longbibliography]{revtex4-1}

\usepackage{graphicx}
\usepackage{amsmath}
\usepackage{amsfonts}
\usepackage{amssymb}
\usepackage{braket}
\usepackage{bm}
\usepackage{multirow}
\usepackage{color}
\usepackage[normalem]{ulem}
\usepackage{mathrsfs}
\usepackage{mathtools}
\usepackage{float}

\makeatletter

\newcommand{\Rmnum}[1]{\expandafter\@slowromancap\romannumeral #1@}
\makeatother

\usepackage[breaklinks]{hyperref}
\hypersetup{colorlinks=true, linkcolor=blue, citecolor=blue, filecolor=blue, urlcolor=blue}

\AtBeginDocument{%
    \newwrite\bibnotes
    \def\bibnotesext{Notes.bib}
    \immediate\openout\bibnotes=\jobname\bibnotesext
    \immediate\write\bibnotes{@CONTROL{REVTEX41Control}}
    \immediate\write\bibnotes{@CONTROL{%
    apsrev41Control,author="08",editor="1",pages="1",title="0",year="1"}}
     \if@filesw
     \immediate\write\@auxout{\string\citation{apsrev41Control}}%
    \fi
}%

\begin{document}

\title{Hilbert-Space Fragmentation from Strict Confinement}

\author{Zhi-Cheng Yang}
\email{zcyang@umd.edu}
\affiliation{Joint Quantum Institute, NIST/University of Maryland, College Park, MD 20742, USA}
\affiliation{Joint Center for Quantum Information and Computer Science, NIST/University of Maryland, College Park, MD 20742, USA}

\author{Fangli Liu}
\affiliation{Joint Quantum Institute, NIST/University of Maryland, College Park, MD 20742, USA}
\affiliation{Joint Center for Quantum Information and Computer Science, NIST/University of Maryland, College Park, MD 20742, USA}

\author{Alexey V. Gorshkov}
\affiliation{Joint Quantum Institute, NIST/University of Maryland, College Park, MD 20742, USA}
\affiliation{Joint Center for Quantum Information and Computer Science, NIST/University of Maryland, College Park, MD 20742, USA}

\author{Thomas Iadecola}
\email{iadecola@iastate.edu}
\affiliation{Department of Physics and Astronomy, Iowa State University, Ames, Iowa 50011, USA}

\date{\today}

\begin{abstract}
We study one-dimensional spin-1/2 models in which strict confinement of Ising domain walls leads to the fragmentation of Hilbert space into exponentially many disconnected subspaces. Whereas most previous works emphasize dipole moment conservation as an essential ingredient for such fragmentation, we instead require two commuting U(1) conserved quantities associated with the total domain-wall number and the total magnetization. The latter arises naturally from the confinement of domain walls. Remarkably, while some connected components of the Hilbert space thermalize, others are integrable by Bethe ansatz. We further demonstrate how this Hilbert-space fragmentation pattern arises perturbatively in the confining limit of $\mathbb{Z}_2$ gauge theory coupled to fermionic matter, leading to a hierarchy of time scales for motion of the fermions.  This model can be realized experimentally in two complementary settings.
\end{abstract}

\maketitle

{\it Introduction.---}Generic nonintegrable quantum many-body systems eventually reach thermal equilibrium under unitary time evolution from initial states having a finite energy density with respect to the Hamiltonian~\cite{d2016quantum}. Such behavior arises in models satisfying the eigenstate thermalization hypothesis (ETH)~\cite{PhysRevA.43.2046, PhysRevE.50.888}, which posits that highly excited eigenstates of generic Hamiltonians at the same energy density are indistinguishable in the thermodynamic limit
as far as local observables are concerned.

Recent experimental and theoretical investigations indicate that ETH in its strongest form can be violated even in nonintegrable systems with translation symmetry. Experiments on Rydberg-atom chains, where persistent revivals in quench dynamics starting from certain initial states are observed~\cite{bernien2017probing}, led to the identification of certain atypical eigenstates that are embedded in an otherwise thermalizing spectrum~\cite{turner2018weak, PhysRevB.98.155134}. Another mechanism leading to ETH violations is dynamical constraints. Fractonic systems, 
where such constraints manifest themselves in the restricted mobility of excitations,
turn out to be natural candidates along this direction~\cite{PhysRevLett.94.040402, castelnovo2012topological, PhysRevA.83.042330, PhysRevB.94.235157}. Mobility restrictions in fractonic systems can be implemented by imposing both charge $(Q)$ and dipole moment $(P)$ conservation~\cite{PhysRevB.95.115139, PhysRevB.96.035119}, providing a simple guiding principle for systematic studies of constrained models. It is shown in Refs.~\cite{PhysRevX.9.021003, khemani2019local, sala2019ergodicity,moudgalya2019thermalization} that these two conservation laws cause the Hilbert space to fracture into disconnected subspaces that are invariant (i.e. closed) under the action of the Hamiltonian; moreover, these invariant subspaces cannot be distinguished by their $(Q, P)$ quantum numbers alone~\cite{rakovszky2019statistical}. This ``fragmentation" of Hilbert space~\cite{PhysRevB.84.115129,PhysRevLett.110.070602,PhysRevLett.123.036403,khemani2019local,sala2019ergodicity,patil2019hilbert, de2019dynamics, hudomal2019quantum} leads to a broad distribution of the eigenstate entanglement entropies within an energy window, violating the strong ETH.

Fractonic systems bear a phenomenological resemblance to
lattice models exhibiting quasiparticle confinement~\cite{pai2019fractons}. One simple example is the one-dimensional (1D) Ising model in transverse and longitudinal magnetic fields, where the latter induces a confining potential for pairs of Ising domain-wall excitations that grows linearly with their separation~\cite{PhysRevD.18.1259, rutkevich2008energy}. Recent studies of confining systems have mainly focused on physics near the ground state, where domain walls and their bound states are well-defined quasiparticles~\cite{PhysRevA.95.023621, kormos2017real, PhysRevB.97.184205, PhysRevLett.122.150601, pai2019fractons, borla2019confined, verdel2019real, lerose2019quasilocalized, PhysRevB.99.180302, PhysRevLett.111.201601, magnifico2019real}. This leaves open the question of the effects of confinement at finite energy density, where there are generically no well-defined quasiparticles.

In this work, we show that Hilbert-space fragmentation (HSF) can arise in models conserving both domain-wall number $(n_{\rm DW})$ and total magnetization ($S^z$). These two commuting U(1) conserved quantities naturally arise from
\textit{strict confinement}, where isolated domain walls cannot move without changing the $S^z$ quantum number, naturally leading to HSF. We exemplify this phenomenon with a 1D spin-1/2 model that features exponentially many invariant subspaces.  These include exponentially many frozen configurations (i.e., subspaces of dimension one), as well as exponentially large subspaces generated by certain ``root configurations" that we enumerate. The pattern of HSF that we find is extremely rich, featuring large subspaces within which the dynamics is thermalizing, as well as others spanning entire $(n_{\rm DW},S^z)$ sectors that are integrable by Bethe ansatz. We further demonstrate how the same HSF pattern arises perturbatively in the extreme confining limit of a 1D $n_{\rm DW}$-conserving spin model that maps exactly onto $\mathbb{Z}_2$ gauge theory coupled to fermionic matter~\cite{iadecola2019quantum,borla2019confined}, which can be realized experimentally using state-of-the-art techniques in cold atoms~\cite{schweizer2019floquet,gorg2019realization}.  We show that HSF gives rise to a complex hierarchy of timescales for quench dynamics that depends crucially on the initial state. Our results thus establish HSF as a mechanism for slow dynamics in gauge theories at finite energy density.

{\it Model.---}
To see how the simultaneous conservation of $S^z$ and $n_{DW}$ gives rise to HSF, we study a simple model:
\begin{equation}
H \!=\! \sum^{L-1}_{i=2} [J 
P_{i-1,i+2}
(\sigma^+_i \sigma^-_{i+1} \ + \  \sigma^-_i \sigma^+_{i+1}) + \Delta_2\,  \sigma^z_i \sigma^z_{i+2}],
\label{eq:model}
\end{equation}
where $P_{i-1,i+2}=1+\sigma^z_{i-1} \sigma^z_{i+2}$ projects out configurations with opposite spins on sites $i-1$ and $i+2$ (see also Ref.~\cite{PhysRevA.95.023621}).
Note that $[H,\sigma^z_{1,L}]=0$, so that we can fix the two edge spins to point down. Adopting the notation $1 \equiv \ \uparrow, 0 \equiv \ \downarrow$ for the local spin states, we see that the kinetic term in Eq.~(\ref{eq:model}) hops a magnon while preserving $n_{\rm DW}$: $0100 \leftrightarrow 0010$, and $1011 \leftrightarrow 1101$. Since the nearest-neighbor Ising interaction couples to the conserved quantity $n_{\rm DW}$, we add a next-nearest-neighbor Ising interaction $\Delta_2$ to make the model more generic (see below). Eq.~(\ref{eq:model}) has two U(1) conserved quantities $(n_{\rm DW}, S^z)$; for our choice of boundary conditions, we have
$n_{\rm DW}=0, 2, \cdots, L-2$, and $S^z=-L+n_{\rm DW}, -L+n_{\rm DW}+2, \cdots, L-n_{\rm DW}-2$ for $n_{\rm DW} \neq 0$. This gives rise to $\sum_{n_{\rm DW}=2}^{L-2} (L-n_{\rm DW})+1 =\frac{L}{2}(\frac{L}{2}-1)+1$ sectors labeled by these two quantum numbers. As we show later [see Eq.~(\ref{eq:confine})], one can think of Hamiltonian~(\ref{eq:model}) as describing an $n_{\rm DW}$-conserving spin-1/2 system in a uniform confining longitudinal field $h\sum_i \sigma^z_i$ in the strict-confinement limit $h\to\infty$.  In this limit, $S^z$ becomes a conserved quantity. Isolated domain walls (``quarks") cannot move without changing $S^z$, costing infinite energy. However, tightly bound pairs of domain walls (magnons, or ``mesons") can move without violating $S^z$ conservation.

\begin{figure}[!t]
\includegraphics[width=.48\textwidth]{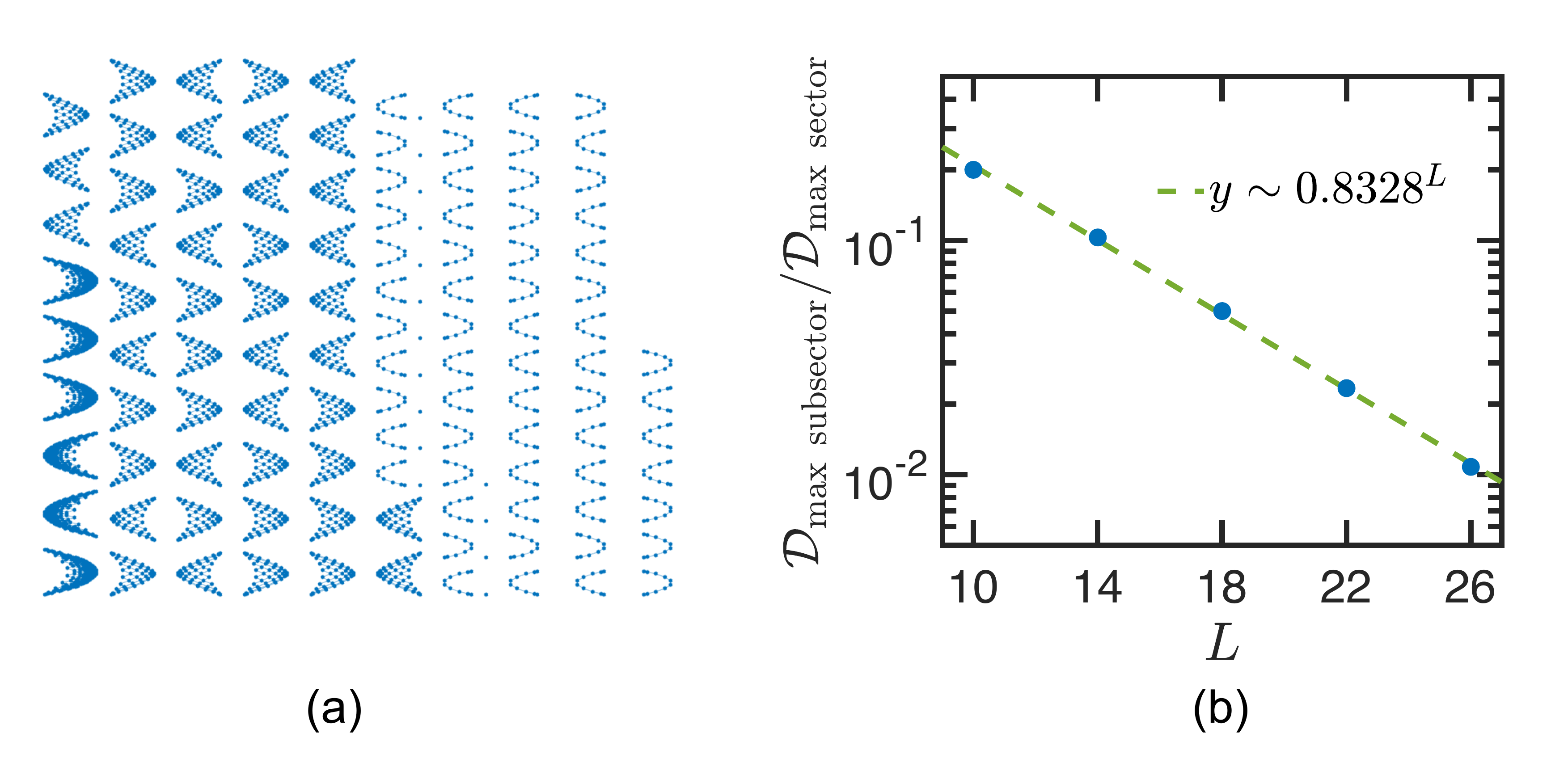}
\caption{(a) Connectivity within the sector $(n_{\rm DW}=8, S^z=-2)$ for $L=18$. This sector has a total Hilbert space dimension of 4410. (b) Ratio of the size of the largest emergent subsector {\it within} the largest $(n_{\rm DW}, S^z)$ sector to that of the entire sector, for different system sizes.}
\label{fig:fragmentation} 
\end{figure}

{\it Strong HSF.---}Naively, one would expect that the Hilbert space of Hamiltonian~(\ref{eq:model}) organizes into $\mathcal{O}(L^2)$ symmetry sectors. In Fig.~\ref{fig:fragmentation}(a), we visualize the symmetry
sector $(n_{\rm DW}=8, S^z=-2)$ as a graph whose nodes are $z$-basis configurations and whose edges correspond to nonzero matrix elements of $H$. We find that the Hilbert space {\it within} this symmetry sector further fractures into many disconnected emergent subsectors (invariant subspaces) of various sizes. In particular, there are isolated nodes in Fig.~\ref{fig:fragmentation}(a), indicating the existence of frozen configurations constituting subsectors of dimension one. In Fig.~\ref{fig:fragmentation}(b) we show that Hamiltonian~(\ref{eq:model}) exhibits {\it strong} HSF
as defined in Ref.~\cite{sala2019ergodicity}: the ratio of the dimension of the largest emergent subsector
{\it within} the largest $(n_{\rm DW}, S^z)$ sector to that of the whole sector decreases exponentially with $L$. This implies that in the thermodynamic limit, even the largest emergent subsector constitutes a vanishing fraction of the full $(n_{\rm DW}, S^z)$ sector. Intriguingly, the same HSF pattern arises in a different context in Ref.~\cite{de2019dynamics}, which studies a disordered fermionic system with strong nearest-neighbor interactions.

We now develop an understanding of the pattern of HSF evident in Fig.~\ref{fig:fragmentation}, starting with the origin of the frozen states in Fig.~\ref{fig:fragmentation}(a). 
As discussed below Eq.~\eqref{eq:model}, the only nonzero off-diagonal matrix elements of $H$ are between configurations differing by the nearest-neighbor exchange of a single magnon. This immediately implies that the kinetic term in Eq.~\eqref{eq:model} annihilates any configuration containing no isolated magnons, and that such configurations are disconnected from all others.  Since an isolated magnon is equivalent to a pair of domain walls occupying neighboring bonds, it follows that any configuration in which no two neighboring bonds host a domain wall is frozen (see Supplemental Materials~\cite{SM}).  This nearest-neighbor exclusion is sometimes called the ``Fibonacci constraint," which also arises in systems of Rydberg atoms with strong interactions~\cite{bernien2017probing}. The number of states satisfying this constraint grows as $\varphi^L$, where $\varphi$ is the golden ratio. Configurations in which \textit{every} bond is occupied by a domain wall (e.g.~$0101\dots$) are also frozen because domain walls are hardcore objects; however, the number of such configurations is independent of system size~\cite{SM}.

Next, we identify a class of root configurations from which each connected subsector can be built. Consider configurations of the following form:
\begin{equation}
0 \ \underbrace{\fbox{frozen state}}_{L-2-2k} \ \underbrace{\fbox{$\displaystyle 0101\cdots01$}}_{2k} \ 0,
\label{eq:root}
\end{equation}
which are constructed by appending a N\'eel state of length $2k$ to the right of any magnon-free frozen state. The two outermost 0's denote the edge spins that remain fixed. Since the N\'eel region contains $k$ magnons, we shall call~(\ref{eq:root}) a ``$k$-magnon state."
One can explicitly check that, although the two constituent subsystems are both frozen, the boundary between them becomes active once they are joined together~\cite{SM}.
To show that any connected subsector can be built from a $k$-magnon state, we first point out an important property in our system which is in stark contrast to spin-1 systems with $(Q, P)$ conservation~\cite{sala2019ergodicity,khemani2019local,khemani2019localization}. 
Whereas these models allow mobile excitations to be contained within a finite domain by constructing appropriate ``shielding regions," there are no such regions in the model \eqref{eq:model}:
an isolated mobile magnon 
can propagate all the way to the boundary of the system. Therefore, the model \eqref{eq:model} does not support spatially separated thermal and nonthermal domains, while fractonic systems do~\cite{sala2019ergodicity,khemani2019local,khemani2019localization}. Using this fact, one can then prove~\cite{SM} that any configuration that is not frozen can be brought into the form~(\ref{eq:root}) by propagating all mobile magnons to the right boundary using Eq.~\eqref{eq:model}.  Therefore, any connected subsector can be built from an appropriate $k$-magnon state.

\begin{figure}[!t]
\includegraphics[width=.5\textwidth]{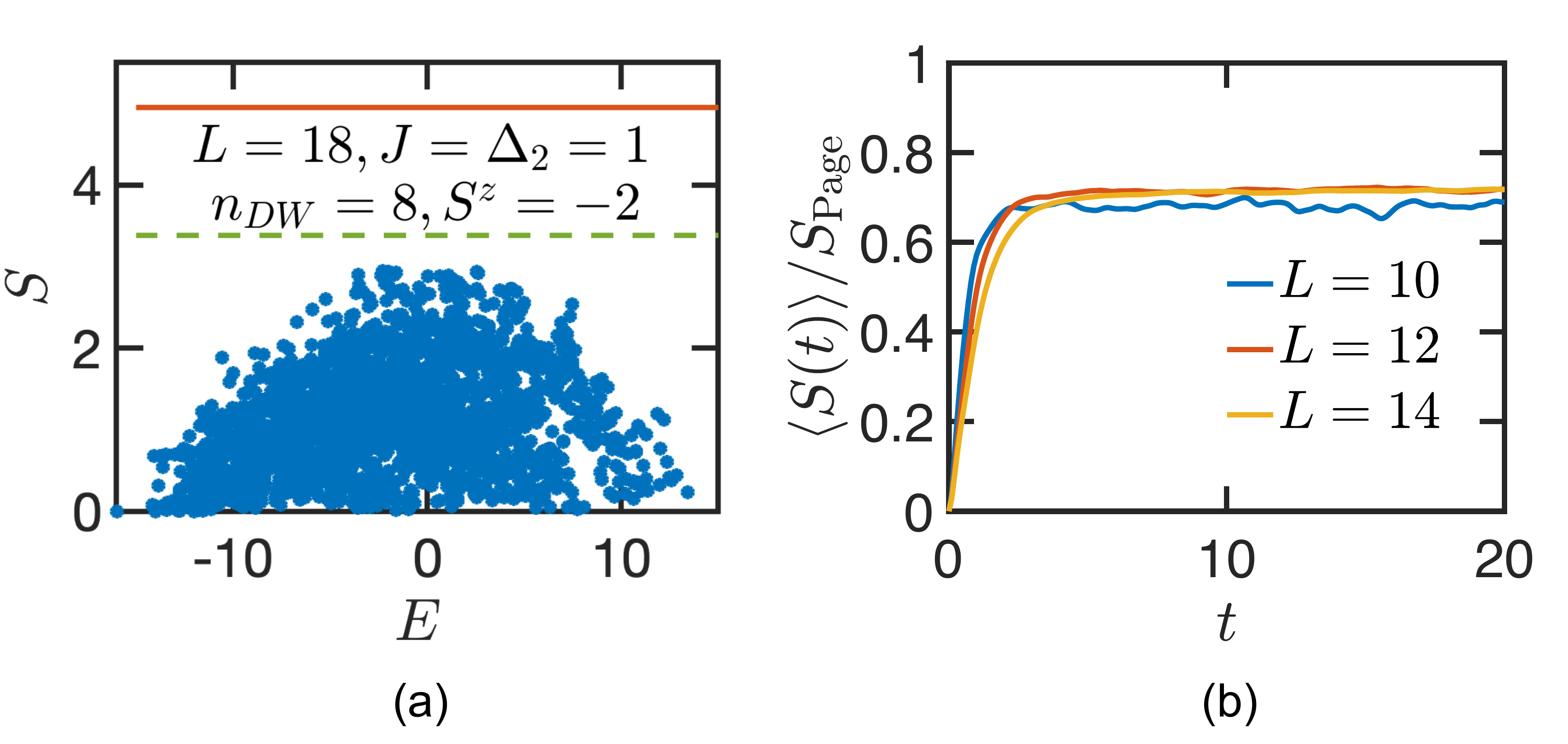}
\caption{(a) Entanglement entropy of the eigenstates within the sector $(n_{\rm DW}=8, S^z=-2)$ under an equi-bipartitioning of the system.
Red line: Page value of the $(n_{\rm DW}, S^z)$ sector; green line: Page value of the largest connected subsector. (b) Entanglement entropy growth (normalized by the Page value) after a quantum quench starting from random product states, averaged over 200 initial states.}
\label{fig:entanglement} 
\end{figure}

{\it Subsector thermalization and integrability.---}The fracturing of the Hilbert space into exponentially many disconnected subsectors indicates that the eigenstates of Hamiltonian~(\ref{eq:model}) strongly violate ETH, as can be diagnosed from the entanglement entropy. In Fig.~\ref{fig:entanglement}(a), we plot the entanglement entropy of the eigenstates within an $(n_{\rm DW}, S^z)$ symmetry sector. There is clearly a broad distribution in the entanglement entropy, even for eigenstates that are close in energy. In particular, the frozen states have exactly zero entanglement entropy although they reside in the middle
of the energy spectrum. Moreover, the maximal value of the entanglement entropy stays far below the ``Page value," i.e., that of a random state in the corresponding $(n_{\rm DW}, S^z)$ sector~\cite{PhysRevLett.71.1291}. The non-thermalizing behavior of the full Hamiltonian also manifests itself in quantum quenches starting from random initial product states that do not belong to any particular symmetry sector. In Fig.~\ref{fig:entanglement}(b), we find that the final entanglement entropy under time evolution only saturates to 70\% of the Page value, confirming that the system does not thermalize under time evolution.

\begin{figure}[!t]
\includegraphics[width=.5\textwidth]{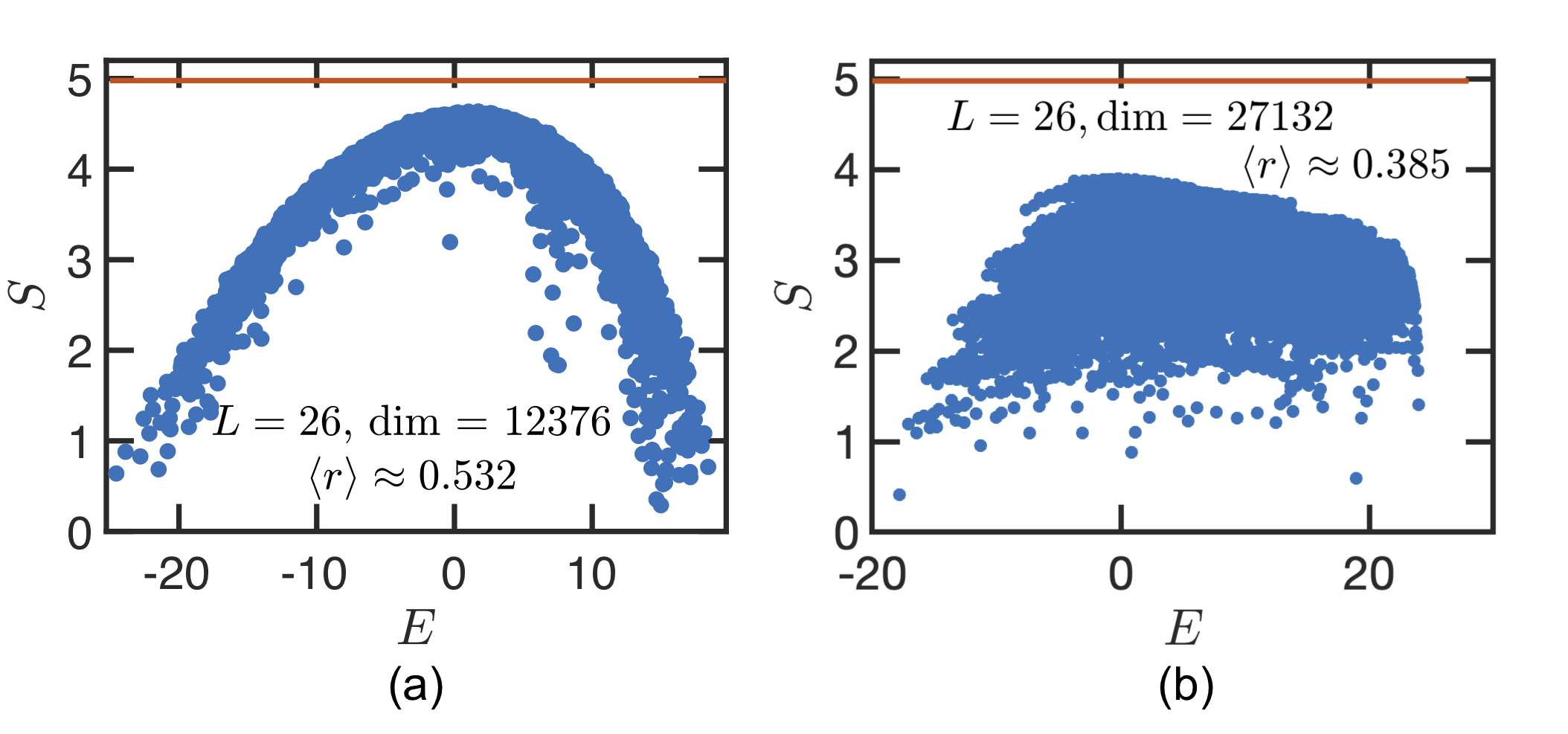}
\caption{(a) Entanglement entropy of eigenstates within an emergent subsector built from the root configuration $0 \fbox{111111000000} \fbox{010101010101} 0 $ for system size $L=26$. This subsector has dimension 12376 and is nonintegrable. (b) Entanglement entropy of eigenstates within an emergent subsector built from the root configuration $0 \fbox{000000000000} \fbox{010101010101} 0 $ for system size $L=26$. This subsector has dimension 27132 and is integrable. Red lines mark the Page value of the corresponding subsector.}
\label{fig:entropy_subsector} 
\end{figure}

The fragmentation of Hilbert space seems to suggest that a more appropriate comparison of the entanglement entropy might be the Page value {\it restricted} to a connected subsector. To this end, we extract the effective Hilbert space dimensions of the left and right halves of the chain $\mathcal{D}_L$ and $\mathcal{D}_R$ within the largest emergent subsector, and then compute the corresponding Page value using the exact formula: $\sum_{k=n+1}^{mn}\frac{1}{k}-\frac{m-1}{2n}$, where $m={\rm min}[\mathcal{D}_L, \mathcal{D}_R]$, and $n={\rm max}[\mathcal{D}_L, \mathcal{D}_R]$~\cite{PhysRevLett.71.1291}. As shown in Fig.~\ref{fig:entanglement}(a) (green dashed line), the maximal eigenstate entanglement entropy is close to the Page value restricted to the largest subsector. This strongly indicates that the system thermalizes \textit{within} each invariant subspace~\cite{moudgalya2019thermalization}. Testing this scenario numerically requires larger system sizes with bigger subsector dimensions. Fortunately, armed with the knowledge of the root configurations~(\ref{eq:root}), one can directly construct the projection of Hamiltonian \eqref{eq:model} into an arbitrary emergent subsector. In Fig.~\ref{fig:entropy_subsector}(a), we show the entanglement entropy for eigenstates within a connected subsector built from the root configuration $0 \fbox{111111000000} \fbox{010101010101} 0 $. It is clear
that the eigenstate entanglement entropy within this subsector forms a narrow ETH-like band, with maximal value close to the subspace-restricted Page value. Moreover, we compute the average energy level spacing ratio for the eigenenergies of the projected Hamiltonian: $r_i = {\rm min}\{\delta_i, \delta_{i+1}\}/{\rm max}\{\delta_i, \delta_{i+1}\}$, where $\delta_i = E_i-E_{i+1}$ is the gap between adjacent energy levels~\cite{PhysRevB.82.174411}. We find $\langle r \rangle \approx 0.532$, consistent with the Gaussian orthogonal ensemble in random matrix theory~\cite{PhysRevB.82.174411}. Taken together, these facts suggest that there is indeed a notion of \textit{subsector thermalization} in the present model. In the absence of $\Delta_2$ in Eq.~(\ref{eq:model}), we numerically find that the spectral properties strongly deviate from nonintegrability, which confirms the necessity of including a nonzero $\Delta_2$.


At this point, it may seem that {\it all} sufficiently large connected subsectors at finite energy density thermalize when considered separately. However, as we now show, this is not the case. Consider the sequence of symmetry sectors $(n_{\rm DW}=2k, S^z=-L+2k)$, which have the smallest possible $S^z$ for a given $n_{\rm DW}$. These sectors can be generated from root configurations $0 \fbox{$\displaystyle 00\cdots0$} \fbox{$\displaystyle 0101\cdots01$} 0$ and are in fact fully connected, i.e, they do not fracture into subsectors. The projection of Hamiltonian~(\ref{eq:model}) into these symmetry sectors yields a constrained XXZ model in which neighboring up spins are separated by at least two sites~\cite{alcaraz1999exactly, borla2019confined}. For Hamiltonian~\eqref{eq:model} this constraint is automatically satisfied within these symmetry sectors, since bringing two up spins next to one another annihilates a pair of domain walls, which is forbidden by $n_{\rm DW}$ conservation. Remarkably, the constrained XXZ model, although interacting, is exactly solvable via Bethe ansatz, and hence integrable~\cite{alcaraz1999exactly}. This is also seen numerically in Fig.~\ref{fig:entropy_subsector}(b), where the entanglement entropy does not form an ETH-like band, and where
$\langle r \rangle \approx 0.385$ indicates Poissonian energy-level statistics characteristic of integrability~\cite{PhysRevB.82.174411}. Notice from Fig.~\ref{fig:entropy_subsector}(b) that, although these sectors are integrable, they reside within the same energy window as the nonintegrable subsectors.

\begin{figure}[!t]
\includegraphics[width=.5\textwidth]{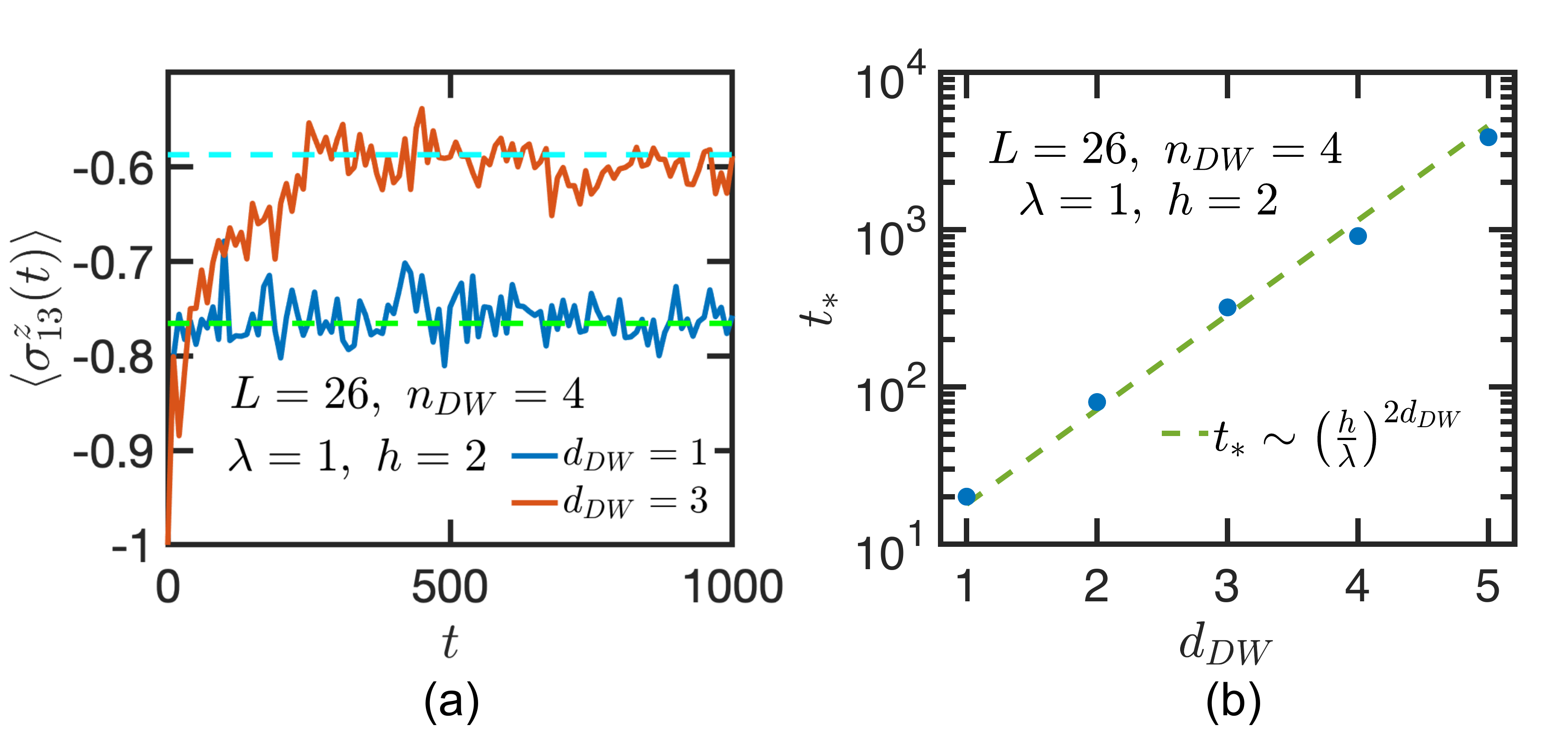}
\caption{(a) Expectation value of the spin in the middle of the chain under time evolution with Hamiltonian~(\ref{eq:confine}), starting from initial configurations with two pairs of domain walls: $00\cdots011\cdots100\cdots011\cdots100\cdots0$. The dashed lines mark the diagonal ensemble average values: $\langle \sigma^z_{13} \rangle_{\rm diag} = \sum_n \langle n| \sigma^z_{13}|n\rangle |c_n|^2$, where $|n\rangle$ denotes the eigenstate of the Hamiltonian and $c_n =\langle n|\psi_0\rangle$ is the overlap between the initial state and each eigenstate. (b) Scaling of the saturation timescale $t_*$ as a function of $d_{DW}$.}
\label{fig:timescale} 
\end{figure}

{\it HSF in gauge theory.---} We now show how the pattern of HSF described above arises in the strict-confinement limit of $\mathbb Z_2$ gauge theory coupled to fermionic matter, and study its breakdown as the strict-confinement limit is relaxed. We first demonstrate that the pattern of HSF observed in Hamiltonian~(\ref{eq:model}) naturally arises in the $n_{\rm DW}$-conserving model~\cite{iadecola2019quantum, borla2019confined}
\begin{equation}
H_{\mathbb{Z}_2} = \sum_i [\lambda (\sigma^x_i - \sigma^z_{i-1} \sigma^x_i \sigma^z_{i+1}) + h \sigma^z_i].
\label{eq:confine}
\end{equation}
As shown in Ref.~\cite{borla2019confined}, this model maps exactly onto $\mathbb Z_2$ gauge theory coupled to spinless fermions in 1D, where the Ising domain-wall number operator in the spin model is reinterpreted as the fermion number operator in the gauge theory.  With this in mind, we will henceforth use the terms ``domain wall" and ``fermion" interchangeably.
The kinetic term in Eq.~\eqref{eq:confine} induces nearest-neighbor hopping of domain walls, while
the longitudinal field introduces a linearly confining potential. This model can be realized experimentally in two complementary settings. The spin model \eqref{eq:confine} can be realized by Floquet engineering in periodically-driven transverse-field Ising chains~\cite{PhysRevB.92.125107, PhysRevLett.119.123601}, while the gauge theory itself can be realized in ultracold atomic gases~\cite{barbiero2019coupling}. Experimental steps toward the latter have already been reported in Refs.~\cite{schweizer2019floquet,gorg2019realization}.

To understand the effect of confinement in Eq.~(\ref{eq:confine}), we work in the limit $h \gg \lambda$. At $h=\infty$, the energy spectrum of Hamiltonian~(\ref{eq:confine}) becomes highly degenerate, with each $S^z$ sector forming a degenerate manifold. The dynamics at $h=\infty$ is trivial; the leading nontrivial behavior is 
determined by performing degenerate perturbation theory in $\lambda/h$. Formally, this is carried out by a Schrieffer-Wolff transformation~\cite{SM}, which yields an effective Hamiltonian $H_{\rm eff}=\sum_{n}H^{(n)}_{\rm eff}$, where $H^{(n)}_{\rm eff}$ is of order $(\lambda/h)^n$ and conserves $n_{\rm DW}$ and $S^z$ by construction. Strictly speaking this analysis is valid up to an order $n_*\sim h/\lambda$ (up to logarithmic corrections), which sets an exponentially long prethermal time scale $\sim e^{c n_*}$ for some constant $c$~\cite{abanin2017rigorous}.

The leading contribution in perturbation theory appears at second order~\cite{SM}:
\begin{equation}
\label{eq:Heff2}
H_{\rm eff}^{(2)} \!=\!\! \frac{\lambda^2}{h} \sum_i [
\sigma^z_{i-1}
P_{i-1,i+2}
(\sigma^+_i \sigma^-_{i+1} + {\rm H.c.}) -\sigma^z_{i-1}\sigma^z_i \sigma^z_{i+1}].
\end{equation}
The kinetic term in Eq.~\eqref{eq:Heff2} coincides with that of Eq.~\eqref{eq:model} up to a configuration-dependent local sign due to the extra factor of $\sigma^z_{i-1}$; this only affects the signs of certain matrix elements, so that Eqs.~\eqref{eq:Heff2} and \eqref{eq:model} exhibit the same pattern of HSF.  Moreover, although Eq.~\eqref{eq:Heff2} sports a three- rather than a two-body interaction, this has no effect on the (non)integrability of the various (sub)sectors. In the integrable sectors, the spin between any two up spins must point down by $(n_{\rm DW},S^z)$ conservation. The three-body interaction in $H_{\rm eff}^{(2)}$ thus reduces (up to a constant shift) to $\Delta_2$ upon setting $\sigma^z_i=-1$ in $\sigma^z_{i-1}\sigma^z_i\sigma^z_{i+1}$. Moreover,
the nonintegrable subsectors remain nonintegrable regardless of whether $\Delta_2$ or the three-body term is used. In~\cite{SM}, we numerically verify the above claims by reproducing Figs.~\ref{fig:entanglement} and \ref{fig:entropy_subsector} using $H_{\rm eff}^{(2)}$.

Corrections to the pattern of HSF discussed so far arise for $n>2$, where further-neighbor domain-wall hopping processes appear~\cite{SM}. Such processes reduce the strong HSF of Eq.~\eqref{eq:Heff2} to \textit{weak} HSF, defined in Ref.~\cite{sala2019ergodicity}; in particular, each $(n_{\rm DW},S^z)$ sector collapses into an exponentially large connected cluster that remains disconnected from a set of exponentially many frozen configurations.  The base of the exponential number of such frozen configurations depends on the order in perturbation theory being considered; for example, at $n=4$ the number of frozen states grows as $1.466^L$~\cite{SM}. One can show that a pair of domain walls separated by a distance $d_{\rm DW}$ becomes mobile at order $(\lambda/h)^{2d_{\rm DW}}$ in perturbation theory~\cite{pai2019fractons, SM}. Thus, a configuration containing two domain walls with $d_{\rm DW}>1$, which is frozen at second order, remains frozen for any $n<2d_{\rm DW}$.  Frozen configurations with $n_{\rm DW}>2$ unfreeze at order $n=\text{min}(d_{\rm DW})$, where the minimum is taken over all pairs of domain walls.

The preceding considerations indicate that the thermalization time when evolving with Eq.~\eqref{eq:confine} from a configuration with minimum domain-wall separation $d_{\rm DW}$ will be lower-bounded by a timescale $t_*\sim (h/\lambda)^{2d_{\rm DW}}$. In Fig.~\ref{fig:timescale}(a), we show the evolution under Eq.~\eqref{eq:confine} of
$\langle \sigma^z_{L/2}(t)\rangle$, 
starting from initial configurations with two well-separated pairs of domain walls: $00\cdots0\underbrace{11\cdots1}_{d_{\rm DW}}00\cdots0\underbrace{11\cdots1}_{d_{\rm DW}}00\cdots0$. Indeed, we find that, even for reasonably small $h/\lambda=2$, the timescale for the local observable to saturate to the diagonal-ensemble value~\cite{rigol2008thermalization} expected at late times is longer for initial states with a larger $d_{\rm DW}$. Scaling analysis of this timescale is also in agreement with the prediction $t_*\sim (h/\lambda)^{2d_{\rm DW}}$, as shown in Fig.~\ref{fig:timescale}(b). We thus find that the above reasoning provides a basis to estimate relaxation timescales in the confining limit of the gauge-theory model \eqref{eq:confine} and the correlations between these timescales and the initial state. Deeper investigations of these timescales could be carried out in experimental realizations of the model~\eqref{eq:confine}.

{\it Conclusion.---}In this work, we demonstrate that HSF naturally arises in lattice models exhibiting strict confinement. 
We uncover a highly unusual feature in the models we study, namely the coexistence of nonintegrable emergent subsectors with Bethe-ansatz integrable fully connected symmetry sectors. This work also elucidates the role of HSF in determining the hierarchy of relaxation timescales in the confining phases of lattice gauge theories and related spin models in 1D, paving the way for experimental tests of these ideas in emerging quantum simulation platforms.
These ideas can be generalized to higher dimensions, e.g., by allowing magnons to hop only if they remain isolated. We leave this for future work.

We thank Umberto Borla, Yang-Zhi Chou, Fabian Grusdt, Sergej Moroz, Sanjay Moudgalya, Abhinav Prem, Ruben Verresen, and Tibor Rakovszky for useful discussions. This work is supported by AFOSR FA9550-16-1-0323 (Z.-C. Y.), Iowa State University startup funds (T.I.), NSF PFCQC program, DoE BES QIS program (award No. DE-SC0019449), DoE ASCR FAR-QC (award No. DE-SC0020312), DoE ASCR Quantum Testbed Pathfinder program (award No. DE-SC0019040), AFOSR, ARO MURI, ARL CDQI, and NSF PFC at JQI (Z.-C. Y., F. L., and A. V. G.).
 

\bibliography{reference}


\newpage
\onecolumngrid
\appendix

\subsection*{Supplemental Material for ``Hilbert-Space Fragmentation From Strict Confinement"}

\section{Counting of frozen states}

In this section, we prove that Hamiltonian~(\ref{eq:model}) harbors exponentially many exactly frozen eigenstates in its spectrum. The proof follows from an inductive method analogous to Ref.~\cite{khemani2019local}.
Starting from $L=4$, it is easy to enumerate explicitly that there are 12 frozen states. Suppose we have a frozen state of size $L$ and we would like to increase its size by one, going from $L$ to $L+1$, such that the longer chain remains frozen. Since the kinetic term in Hamiltonian~(\ref{eq:model}) involves at most four spins, the new dynamics introduced by the added spin only depends on the last three spins close to the edge of the original chain. For example, if the last three spins of the original chain are 000, then the added spin can be either 0 or 1, and the new state of size $L+1$ remains frozen. However, if the last three spins are 001 instead, the added spin must be 1 otherwise the new state becomes active. It is straightforward to enumerate all $2^3=8$ possibilities of the last three spins' configurations and the allowed state(s) of the added spin, which we list below:
\begin{center}
\begin{tabular}{c|c}
spin configuration of the last three sites of system size $L$  \quad  &  \quad added spin state can be  \\
\hline
000 & 0 or 1  \\
001 & 1 \\
010 & 1 \\
011 & 0 or 1 \\
100 & 0 or 1 \\
101 & 0 \\
110 & 0 \\
111 & 0 or 1 \\
\hline
\end{tabular}
\label{table:induction}
\end{center}

\begin{figure}[!t]
\includegraphics[width=.45\textwidth]{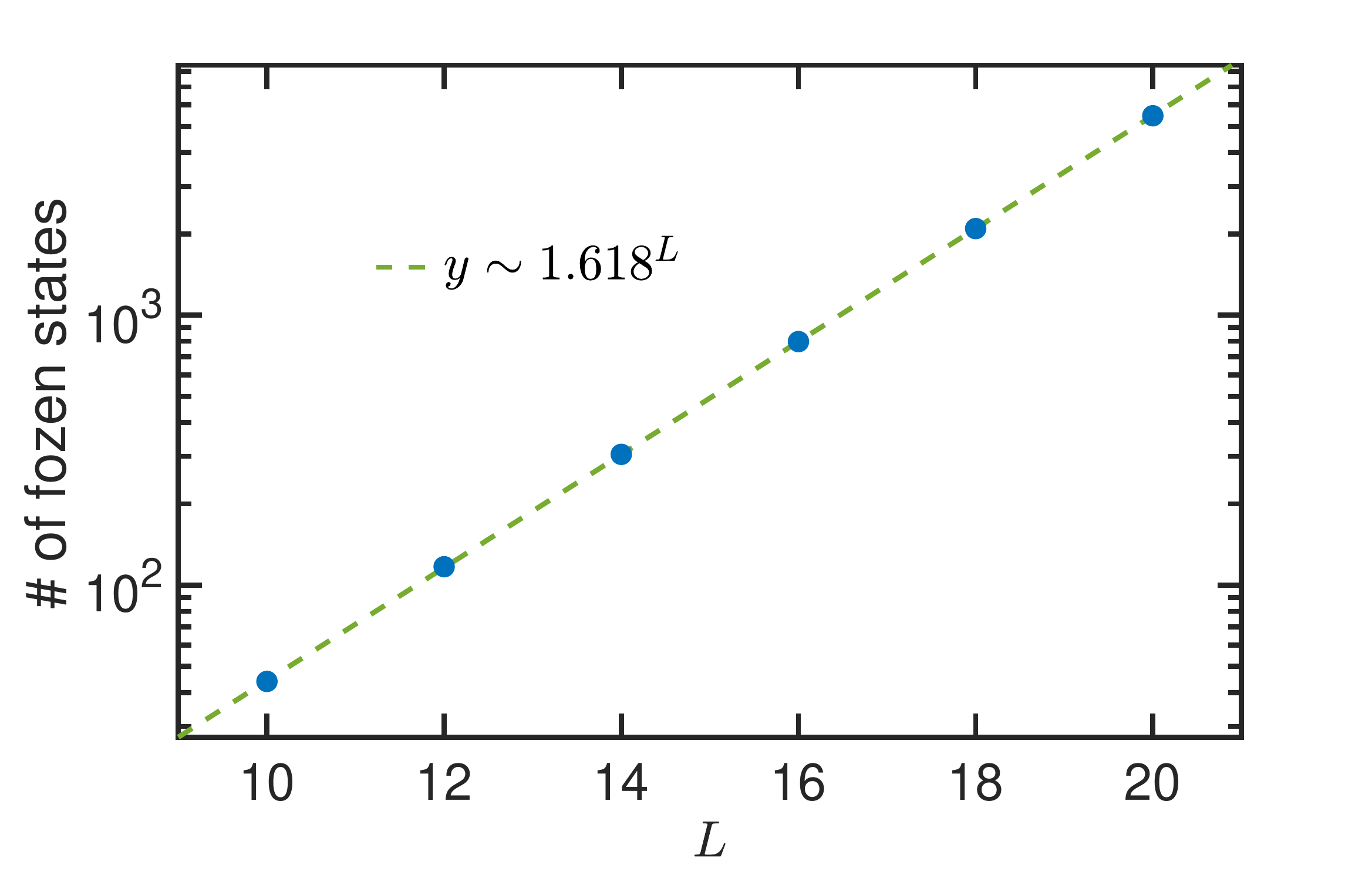}
\caption{Scaling of the total number of frozen states as a function of the system size. The result agrees with the scaling form $y\sim 1.618^L$.}
\label{fig:frozen} 
\end{figure}

Let $N_{abc}(L)$ be the number of frozen states in a system of size $L$ with spin configurations of the last three sites being $abc$. Then $N_{abc}(L+1)$ can be obtained from $N_{abc}(L)$ using Table~\ref{table:induction} as following:
\begin{equation}
\label{eq:induction}
\begin{pmatrix}
N_{000} \\
N_{001} \\
N_{010} \\
N_{011} \\
N_{100} \\
N_{101} \\
N_{110} \\
N_{111} \\
\end{pmatrix}
_{L+1}  = \quad
\begin{pmatrix}
1 & 0 & 0 & 0 & 1 & 0 & 0 & 0  \\
1 & 0 & 0 & 0 & 1 & 0 & 0 & 0  \\
0 & 0 & 0 & 0 & 0 & 1 & 0 & 0  \\
0 & 1 & 0 & 0 & 0 & 0 & 0 & 0  \\
0 & 0 & 0 & 0 & 0 & 0 & 1 & 0  \\
0 & 0 & 1 & 0 & 0 & 0 & 0 & 0  \\
0 & 0 & 0 & 1 & 0 & 0 & 0 & 1  \\
0 & 0 & 0 & 1 & 0 & 0 & 0 & 1  \\
\end{pmatrix}
\begin{pmatrix}
N_{000} \\
N_{001} \\
N_{010} \\
N_{011} \\
N_{100} \\
N_{101} \\
N_{110} \\
N_{111} \\
\end{pmatrix}
_{L}.
\end{equation}
This matrix can be diagonalized to obtain all of its eigenvalues and eigenvectors, which, combined with the initial value $N_{abc}(4)$, can be used to calculate exactly the number of frozen states at arbitrary $L$. However, the asymptotic behavior in the large $L$ limit is controlled by the largest eigenvalue of this matrix $\alpha$, and the number of frozen states goes as $\sim |\alpha|^L$. In this case, we find $\alpha=\frac{1+\sqrt{5}}{2}\equiv \varphi \approx 1.618^L$, which coincides with the asymptotic behavior of the Fibonacci sequence. In Fig.~\ref{fig:frozen}, we check this scaling form numerically and find good agreement.

As explained in the main text, there is indeed an emergent ``Fibonacci constraint" in the frozen subspaces, namely, there cannot be two adjacent domain walls. In the present case, there is one exception to this constraint, which is the N\'eel state $\cdots 010101 \cdots$. Nevertheless, one can see from Eq.~(\ref{eq:induction}) that $N_{010}$ and $N_{101}$ form an independent block, and hence are not important in the asymptotics. Indeed, we find that the corresponding eigenvector of $\varphi$ has zero amplitude on these two components. Therefore, one can ignore the N\'eel configurations as far as only the asymptotics are concerned.

\section{Proof of the existence of ``$k$-magnon state" in each emergent subsector}

We point out in the main text that each emergent subsector can be constructed from the $k$-magnon root state of the following form:
\begin{equation}
0 \ \underbrace{\fbox{frozen state}}_{L-2-2k} \ \underbrace{\fbox{$\displaystyle 0101\cdots01$}}_{2k} \ 0,
\end{equation}
where we append a N\'eel state of length $2k$ to the right of any magnon-free frozen state. By construction, the two subsystems are both inert by themselves. However, the boundary between the two regions will become active. At the boundary of the two regions, the only possible configurations are $00|0101, 11|0101$, or $10|0101$ (by definition $01|01$ cannot be the boundary), and one can see the in any case the boundary contains mobile magnons.

We now prove that each connected subsector contains a $k$-magnon root state of this form. In other words, any configuration that is not frozen can be brought to a $k$-magnon state under Hamiltonian~(\ref{eq:model}). We start by showing the following fact in our model: an isolated mobile magnon inserted in the system can tunnel through the entire system. That is to say, there is no ``shielding region" that can localize a mobile magnon within a certain spatial region, which is in stark contrast to previously studied spin-1 models with $(Q, P)$ conservation.

Consider a single mobile magnon of the form 0100 or 1011 embedded in the system. Consider the configuration of its two neighboring spins to the right (the left can be analyzed in a symmetric way). The two neighboring spins can be 01, 10, 00, or 11. Let us inspect what happens after the mobile magnon moves 1 step:
\begin{eqnarray}
0100 \ | \ 01      \quad        &\rightarrow&     \quad  0\underbrace{010 \ | \ 0}1     \nonumber \\
01\underbrace{00 \ | \ 10}      \quad        &\rightarrow&     \quad  0010 \ | \ 10     \nonumber  \\
0100 \ | \ 00     \quad        &\rightarrow&     \quad  0\underbrace{010 \ | \ 0}0    \nonumber  \\
0100 \ | \ 11      \quad        &\rightarrow&     \quad  00\underbrace{10 \ | \ 11},     \nonumber  
\end{eqnarray}
and
\begin{eqnarray}
10\underbrace{11 \ | \ 01}      \quad        &\rightarrow&     \quad  1101\ | \ 01     \nonumber \\
1011 \ | \ 10      \quad        &\rightarrow&     \quad  1\underbrace{101\ | \ 1}0     \nonumber \\
1011 \ | \ 00      \quad        &\rightarrow&     \quad  11\underbrace{01\ | \ 00}     \nonumber \\
1011 \ | \ 11      \quad        &\rightarrow&     \quad  1\underbrace{101\ | \ 1}1.     \nonumber \\
\end{eqnarray}
In each case above, we denote in brackets the new mobile region that emerges at the boundary of the original mobile region and its neighboring sites. It is thus obvious that, in any case, there will always be new active regions induced by embedding a single mobile magnon into the system. By carrying out the above analysis iteratively, one can demonstrate that this single active magnon can propagate all the way to the right boundary. When the magnon reaches the boundary, since the boundary spin at the right edge is fixed to be zero, the only possible scenarios are the 2nd and 3rd lines in the above processes. For each of these cases, we can check that it can be brought into the form of a $k$-magnon state:
\begin{eqnarray}
01001|0    \quad  &\rightarrow&  \quad  00101|0  \nonumber  \\
01000|0      \quad  &\rightarrow & \quad   00100|0  \quad  \rightarrow \quad 00010|0 \quad \rightarrow \quad 00001|0 \nonumber  \\
10111|0    \quad  & \rightarrow&  \quad 11011|0  \quad \rightarrow \quad 11101|0  \nonumber \\
10110|0  \quad   &\rightarrow&  \quad 11010|0  \quad  \rightarrow  \quad 11001|0  \nonumber 
\end{eqnarray}
Thus, we have shown that the $k$-magnon state exists in each connected subsector, and each subsector can also be constructed using the $k$-magnon state as the root configuration.

\section{Distinctions from fractonic systems and center-of-mass conserving systems}

In this section, we highlight the key distinctions in the mechanism leading to Hilbert-space fragmentation between our model and previously studied fractonic systems~\cite{khemani2019local, sala2019ergodicity} and center-of-mass conserving systems~\cite{moudgalya2019thermalization}.

In fractonic systems, there exist two flavors ($\pm$) of charge excitations that can neutralize into vacuum while preserving the total charge, whereas in our model there is only one type of domain wall excitation. This distinction leads to different allowed local moves in these two models. For example, in fractonic systems, an isolated charge can move at the expense of emitting a dipole, i.e. $0+0 \leftrightarrow +-+$. However, such local moves are completely absent in our model. As a consequence, in fractonic systems mobile excitations can be contained within a finite domain by constructing appropriate ``shielding regions" that essentially cut the chain into disconnected segments. For example, consider a local configuration ++++ embedded in an arbitrary configuration. While the outer two + charges can move by absorbing dipoles, one can easily show that the inner two charges are always immobile. This leads to a {\it spatial coexistence} of thermalizing and frozen regions in the same physical system. However, in our model, an isolated mobile magnon sprinkled into a frozen region can propagate all the way to the boundary of the system. Hence there cannot be spatially separated thermal and non-thermal domains coexisting within a single sample of our model. In fact, it is precisely the particular type of quantum dynamics in our model that enables us to label {\it all} emergent subsectors using a simple class of root configurations, which cannot be simply done in fractonic systems.

Recently, Ref.~\cite{moudgalya2019thermalization} studied a 1D interacting fermion model where particle number and center-of-mass conservation lead to Hilbert-space fragmentation. Despite the apparent similarity, the moves allowed by the kinetic terms in our model are in fact different from Ref.~\cite{moudgalya2019thermalization}. Separating a pair of domain walls in a center-of-mass-preserving manner will necessarily leave a string of either 1's or 0's in between: $010 \rightarrow 01110$ or $101 \rightarrow 10001$, which violates $S^z$ conservation.
This gives rise to completely different {\it pattern} of fragmentation, as well as the structure of the emergent subsectors. In fact, the model in Ref.~\cite{moudgalya2019thermalization} shares many common features with fractonic models as opposed to ours (e.g. the existence of ``shielding regions"), and can be mapped to an extended fractonic model by defining composite degrees of freedom. Moreover, the integrable sectors in Ref.~\cite{moudgalya2019thermalization} are all described by the XX model (or, equivalently, a {\it free fermion} model); in contrast our integrable sectors map to an {\it interacting} model which is not obviously integrable.

Therefore, the mechanism for Hilbert-space fragmentation and its consequence in quantum dynamics in our model are truly distinct from previously studied fractonic systems and center-of-mass conserving systems.

\section{Effective Hamiltonian from Schrieffer-Wolff transformation}
We analyze the effects of confinement in $H_{\mathbb{Z}_2}$ [Eq.~\eqref{eq:confine}] at large $h$ using degenerate perturbation theory in the small parameter $\lambda/h$ based on the Schrieffer-Wolff (SW) transformation~\cite{PhysRev.149.491,bravyi2011schrieffer}.  This is formulated in terms of a unitary transformation
\begin{align}
\label{eq:SWdef}
H_{\rm eff}=e^S\, H\, e^{-S} = H+[S,H]+\frac{1}{2!}[S,[S,H]]+\frac{1}{3!}[S,[S,[S,H]]]+\dots=\sum^{\infty}_{n=0}H^{(n)}_{\rm eff},
\end{align}
where the SW generator $S$ is antiunitary and where $H^{(n)}_{\rm eff}$ is of order $(\lambda/h)^n$.  The choice of $S$ is based on the decomposition
\begin{subequations}
\begin{align}
H&=H_0+V\\
H_0&= h\sum_i \sigma^z_i\\
V&=\lambda\sum_{i}(\sigma^x_i-\sigma^z_{i-1}\sigma^x_{i}\sigma^z_{i+1}).
\end{align}
\end{subequations}
In the local $z$-basis, $H_0$ is diagonal while $V$ is strictly off-diagonal.  In particular, $V$ connects blocks of configurations that differ by a single spin flip, whose energies with respect to $H_0$ differ by $\sim h$ and whose magnetizations $S^z$ differ by 2.  The goal is to choose $S$ such that block-off-diagonal contributions to $H_{\rm eff}$ can be consistently eliminated order by order in $\lambda/h$, so that $[H^{(n)}_{\rm eff},S^z]=0$ for each $n$.

Formally, this can be accomplished by writing
\begin{align}
S = \sum^{\infty}_{n=1}S^{(n)},
\end{align}
where $S^{(n)}$ is of order $(\lambda/h)^n$. Inserting this expression into Eq.~\eqref{eq:SWdef} and grouping terms according to their order in $\lambda/h$ yields
\begin{align}
\label{eq:SWexpansion}
H_{\rm eff}&=H_0\!+\!\left([S^{(1)}\!,H_0]\!+\!V\right)\!+\!\left([S^{(2)}\!,H_0]\!+\![S^{(1)}\!,V]\!+\!\frac{1}{2!}[S^{(1)}\!,[S^{(1)}\!,H_0]]\right)\\
&\quad\!+\!\left([S^{(3)}\!,H_0]\!+\![S^{(2)}\!,V]\!+\!\frac{1}{2!}\left([S^{(1)}\!,[S^{(1)}\!,V]]\!+\![S^{(1)}\!,[S^{(2)}\!,H_0]]\!+\![S^{(2)}\!,[S^{(1)}\!,H_0]]\right)\!+\!\frac{1}{3!}[S^{(1)}\!,[S^{(1)}\!,[S^{(1)}\!,H_0]]]\right)\!+\!\dots.
\nonumber
\end{align}
$S^{(n)}$ is then chosen such that $[S^{(n)},H_0]$ cancels all block-off-diagonal (i.e., non-$S^z$-conserving) terms at order $n$.  This strategy is well-defined and straightforward to automate on a computer (see, e.g., Ref.~\cite{PhysRevB.96.165106}), however it is cumbersome to write out explicitly.

Another (completely equivalent) strategy is to set $S^{(n)}=0$ for $n\geq 2$ and manually project out non-$S^z$-conserving terms at each order. $S^{(1)}$ is still chosen to satisfy $[S^{(1)},H_0]+V=0$, which is accomplished with the choice
\begin{align}
\label{eq:S1def}
\langle\sigma|S^{(1)}|\sigma^\prime\rangle
=\frac{\langle\sigma|V|\sigma^\prime\rangle}{\langle\sigma|H_0|\sigma\rangle-\langle\sigma^\prime|H_0|\sigma^\prime\rangle}.
\end{align}
This gives rise to the leading-order effective Hamiltonian
\begin{subequations}
\begin{align}
H^{(2)}_{\rm eff}&=\mathcal P\left([S^{(1)},V]+\frac{1}{2!}[S^{(1)},[S^{(1)},H_0]]\right)\mathcal P\\
&=\frac{\lambda^2}{h} \sum_i [
\sigma^z_{i-1}
P_{i-1,i+2}
(\sigma^+_i \sigma^-_{i+1} + {\rm H.c.}) -\sigma^z_{i-1}\sigma^z_i \sigma^z_{i+1}],
\end{align}
\end{subequations}
where the first line contains the projection operator $\mathcal P$ that eliminates non-$S^z$-conserving processes and the second line is the result quoted in the main text.

This procedure can be straightforwardly extended to higher orders.  It is readily seen from substituting Eq.~\eqref{eq:S1def} into Eq.~\eqref{eq:SWexpansion} and setting $S^{(n)}=0$ for $n\geq 2$ that $H^{(3)}_{\rm eff}=0$ due to the strictly block-off-diagonal nature of $V$ and hence $S^{(1)}$. (This pattern extends to arbitrary odd orders.) The leading correction to $H^{(2)}_{\rm eff}$ thus arises at fourth order and is given by
\begin{subequations}
\begin{align}
H^{(4)}_{\rm eff}&=\mathcal P\left(\frac{1}{3!}[S^{(1)},[S^{(1)},[S^{(1)},V]]]+\frac{1}{4!}[S^{(1)},[S^{(1)},[S^{(1)},[S^{(1)},H_0]]]]\right)\mathcal P\\
&=\frac{\lambda^4}{2h^3}\sum_i\bigg\{\left(\sigma^z_{i-1}+\sigma^z_{i+3}\right)\left[\frac{3}{2}-\frac{5}{4}\left(\sigma^z_{i-1}\sigma^z_{i+1}+\sigma^z_{i+1}\sigma^z_{i+3}\right)\right]\left(\sigma^+_i\sigma^-_{i+2}+\text{H.c.}\right) \nonumber \\
&\qquad\qquad\qquad  + \left(\sigma^z_{i-1}+\sigma^z_{i+4}\right)\left(\sigma^+_i\sigma^-_{i+2}+\text{H.c.}\right)\left(\sigma^+_{i+1}\sigma^-_{i+3}+\text{H.c.}\right) \label{eq:Heff4} \\
&\qquad\qquad\qquad  -\left(\sigma^z_{i-1}+\sigma^z_{i+4}\right)\left(1-\sigma^z_{i+1}\sigma^z_{i+2}\right)\left(\sigma^+_{i}\sigma^-_{i+1}+\text{H.c.}\right)\left(\sigma^+_{i+2}\sigma^-_{i+3}+\text{H.c.}\right) \bigg\}+\dots, \nonumber
\end{align}
\end{subequations}
where $\dots$ denotes the omission of subleading corrections to matrix elements induced at second order and diagonal terms (i.e., additional interactions) that do not affect Hilbert space connectivity.
The first line of Eq.~\eqref{eq:Heff4} induces matrix elements for processes like $01100 \leftrightarrow 00110$, while the second line leads to processes like $011000\leftrightarrow000110$.  The third line allows for correlated hopping of nearby single magnons, i.e.~$01010\leftrightarrow 00101$.  We thus see that domain walls separated by two sites become mobile at order $(\lambda/h)^4$, as discussed in the main text and in Appendix~\ref{sec:Heff4} below.

\section{Numerical results on the effective Hamiltonian $H_{\rm eff}^{(2)}$}

In this section, we present numerical results demonstrating that the key features of Hamiltonian~(\ref{eq:model}) discussed in the main text can be reproduced by the effective Hamiltonian $H_{\rm eff}^{(2)}$.

\begin{figure}[!t]
\centering
(a)
\includegraphics[width=.42\textwidth]{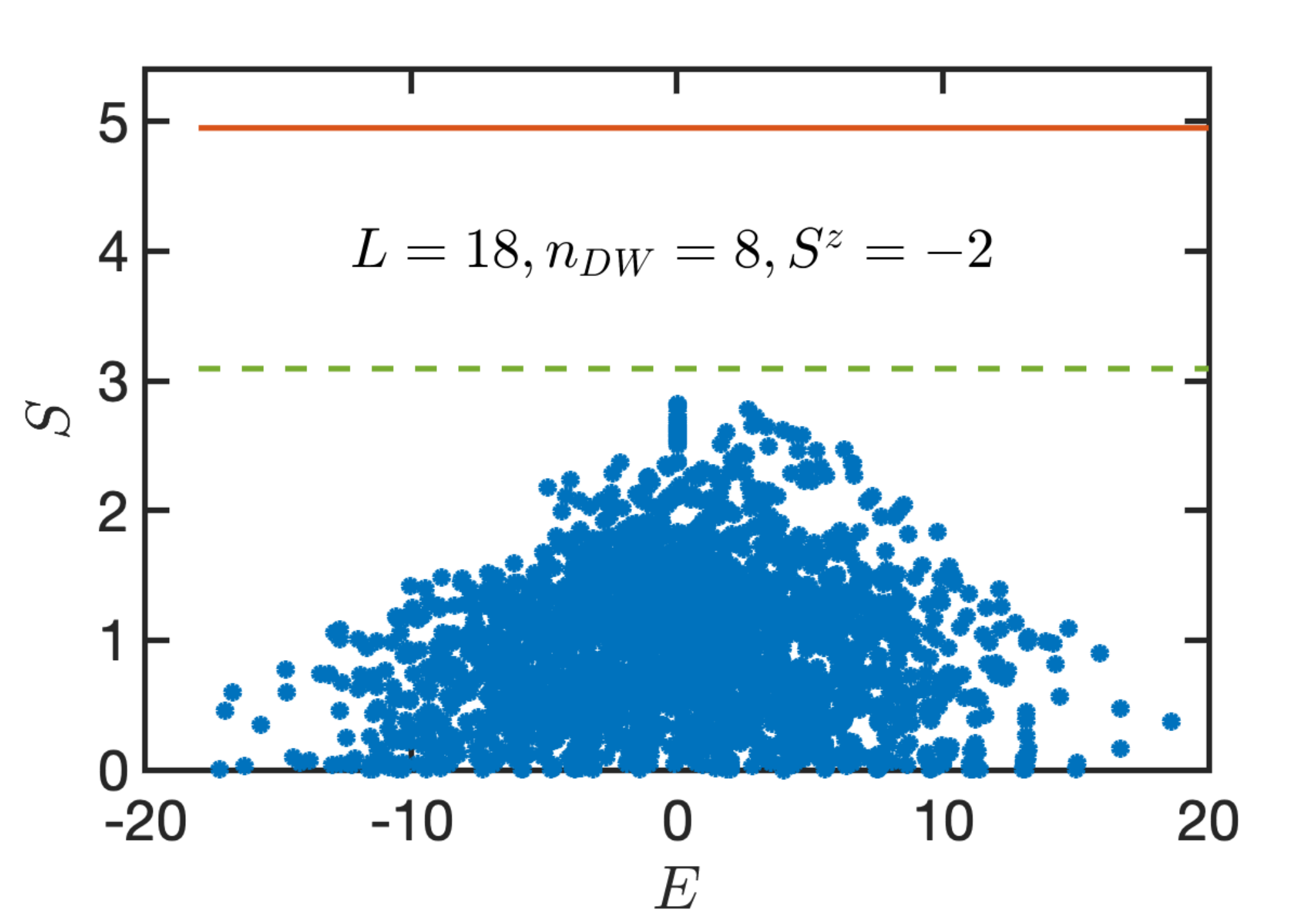}
(b)
\includegraphics[width=.43\textwidth]{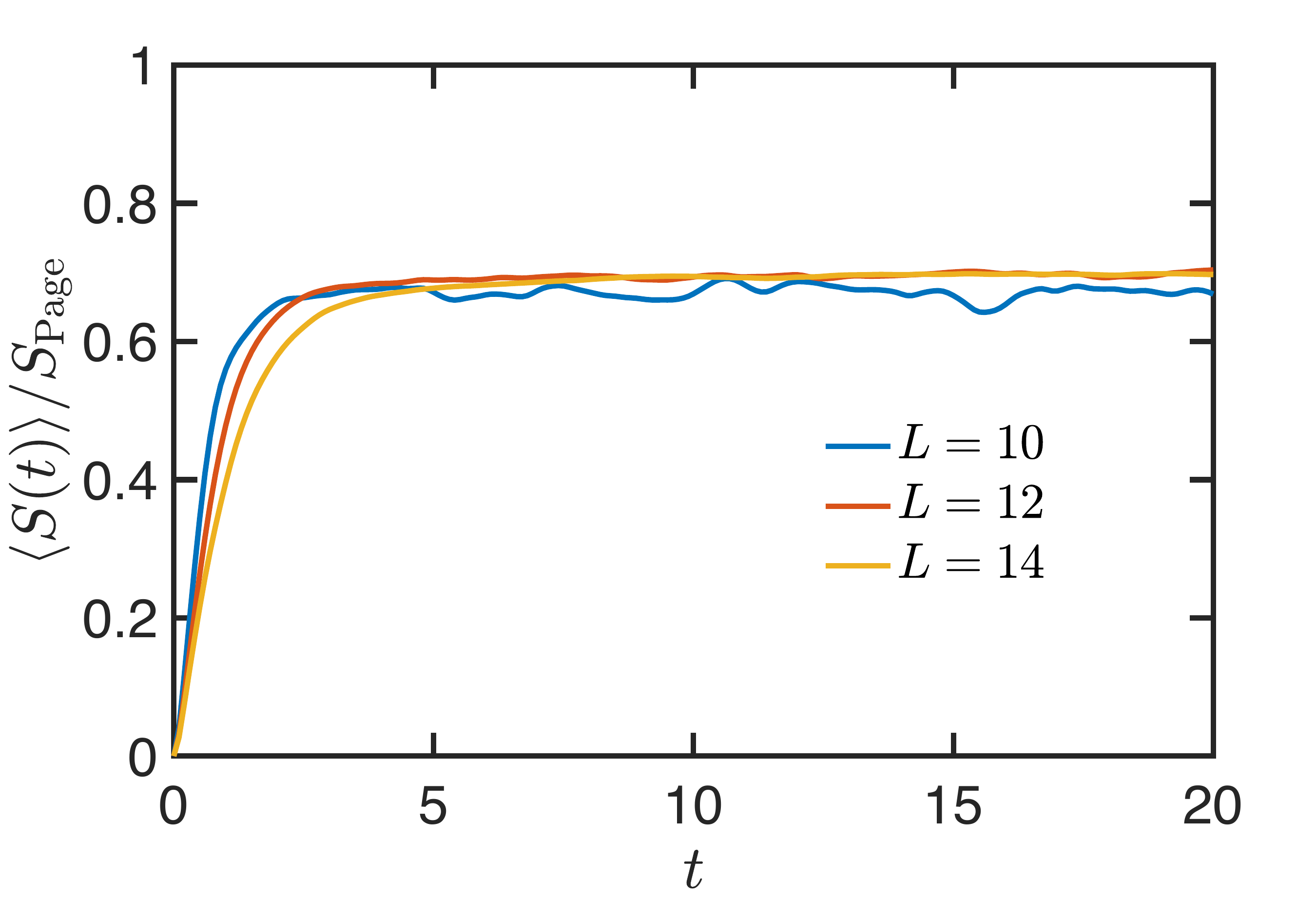}
\\
(c)
\includegraphics[width=.43\textwidth]{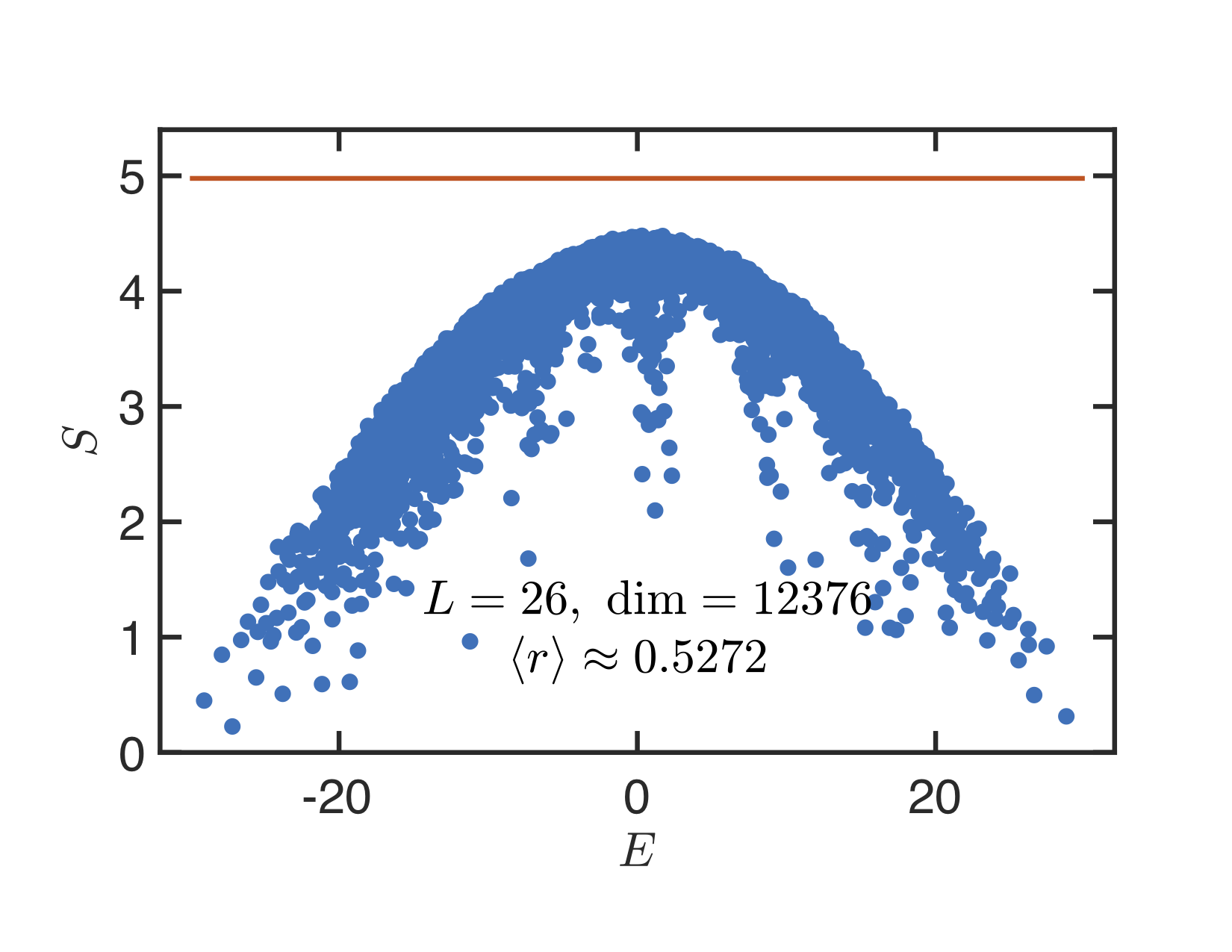}
(d)
\includegraphics[width=.43\textwidth]{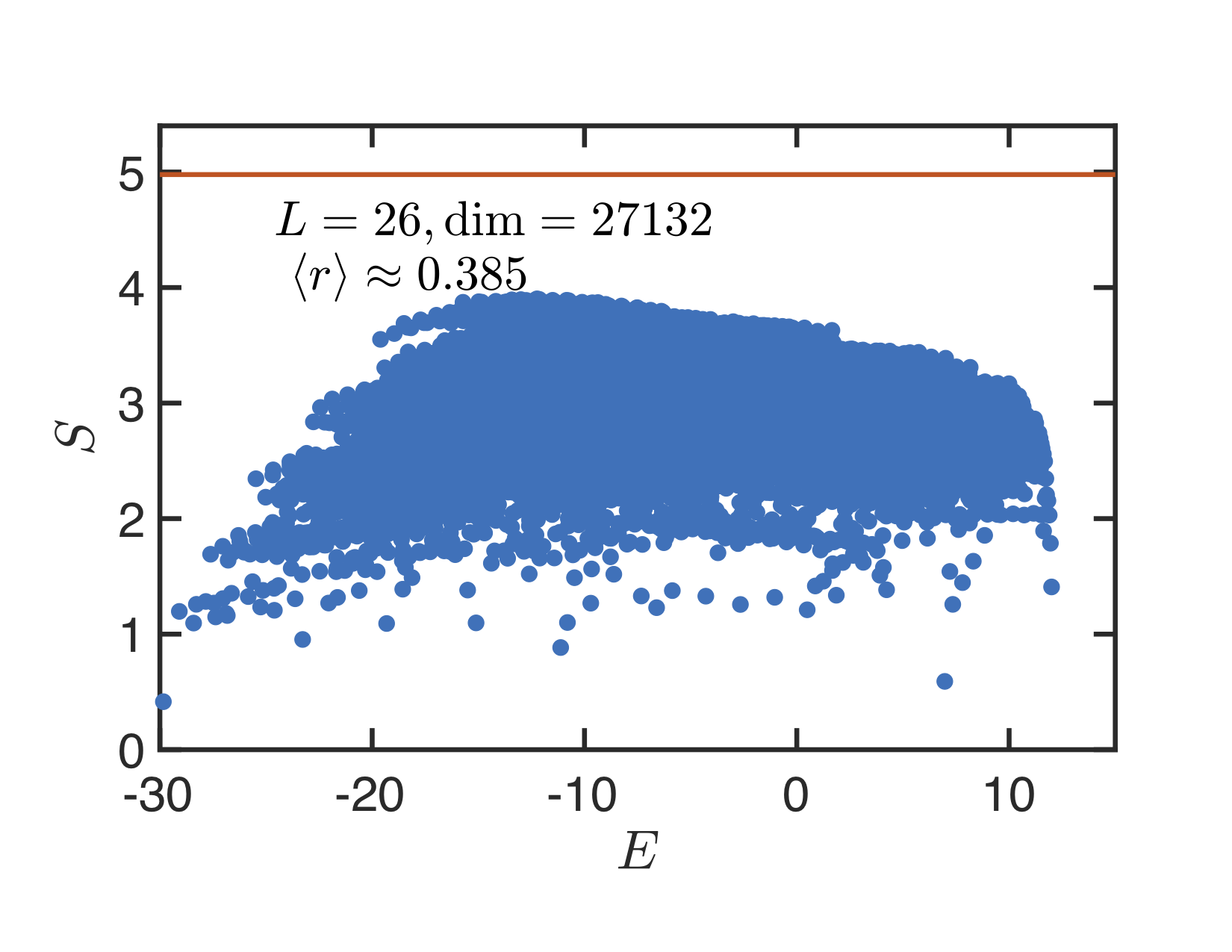}
\caption{(a) Entanglement entropy of the eigenstates of $H_{\rm eff}^{(2)}$ within the sector $(n_{\rm DW}=8, S^z=-2)$ under an equi-bipartitioning of the system in the middle. Orange line: Page value of the $(n_{\rm DW}, S^z)$ sector; green line: Page value of the largest connected subsector. (b) Entanglement entropy growth (normalized by the Page value) after a quantum quench starting from random product states, averaged over 200 initial states. (c) Entanglement entropy of eigenstates within an emergent subsector built from the root configuration $0 \fbox{111111000000} \fbox{010101010101} 0 $ for system size $L=26$. This subsector has dimension 12376 and is nonintegrable. (d) Entanglement entropy of eigenstates within an emergent subsector built from the root configuration $0 \fbox{000000000000} \fbox{010101010101} 0 $ for system size $L=26$. This subsector has dimension 27132 and is integrable. Orange lines mark the Page value of the corresponding subsector.}
\label{fig:Heff}
\end{figure}

In Fig.~\ref{fig:Heff}, we reproduce Figs.~\ref{fig:entanglement} \& \ref{fig:entropy_subsector} shown in the main text, using $H_{\rm eff}^{(2)}$ instead. We have set the overal energy scale in front of $H_{\rm eff}^{(2)}$ to unity.
Indeed, we find good qualitative agreement between Fig.~\ref{fig:Heff} and those in the main text. In Fig.~\ref{fig:Heff}(a), we again find a broad distribution in the entanglement entropy for eigenstates that are close in energy. The maximal value stays far below the Page value for the given symmetry sector. The entanglement entropy evolution after quantum quenches starting from random product states also saturates only to 70\% of the Page value, indicating non-thermal behavior in the long time dynamics under $H_{\rm eff}^{(2)}$ [Fig.~\ref{fig:Heff}(b)].

We further check that the nonintegrable and integrable (sub)sectors remain the same as Hamiltonian~(\ref{eq:model}), despite the slight differences in the sign structure of the kinetic term and the interactions. In Fig.~\ref{fig:Heff}(c), we plot the entanglement entropy of the eigenstates within an emergent subsector. We again find an ETH-like band in the entanglement entropy, with the maximal value close to the subsector-restricted Page value. Moreover, the average energy level spacing ratio gives $\langle r \rangle \approx 0.5272$, which agrees with that of the Gaussian orthogonal ensemble in random matrix theory. This indicates that the same nonintegrable subsector of Hamiltonian~(\ref{eq:model}) in the main text remains nonintegrable under $H_{\rm eff}^{(2)}$. As we have also explained in the main text, when projected to the integrable sectors, $H_{\rm eff}^{(2)}$ once again reduces to a constrained XXZ model which is integrable. In Fig.~\ref{fig:Heff}(d), we plot the entanglement entropy of eigenstates within an integrable sector of Hamiltonian~(\ref{eq:model}). We see that the behavior strongly deviates from that of ETH, and the average energy level spacing ratio yields $\langle r \rangle \approx 0.385$, indicating a Poisson distributed energy spectrum. Therefore, we conclude that the key features of Hamiltonian~(\ref{eq:model}) are indeed captured by Hamiltonian~(\ref{eq:confine}) in the confining limit.

\section{$H_{\rm eff}^{(4)}$ and ``Narayana constraint"}
\label{sec:Heff4}

We now briefly examine the Hilbert space structure of the effective Hamiltonian at fourth-order $H_{\rm eff}^{(4)}$. Under $H_{\rm eff}^{(4)}$, pairs of domain walls separated by distance two become mobile. We find that there are still exponentially many frozen states in the spectrum. However, other than these frozen states, each $(n_{\rm DW}, S^z)$ sector becomes fully connected. Therefore, in this case, we no longer have Hilbert space fragmentation. Instead, we now have exponentially many ``scar" states with exactly zero entanglement entropy embedded in the spectrum.

\begin{figure}[!t]
\includegraphics[width=.45\textwidth]{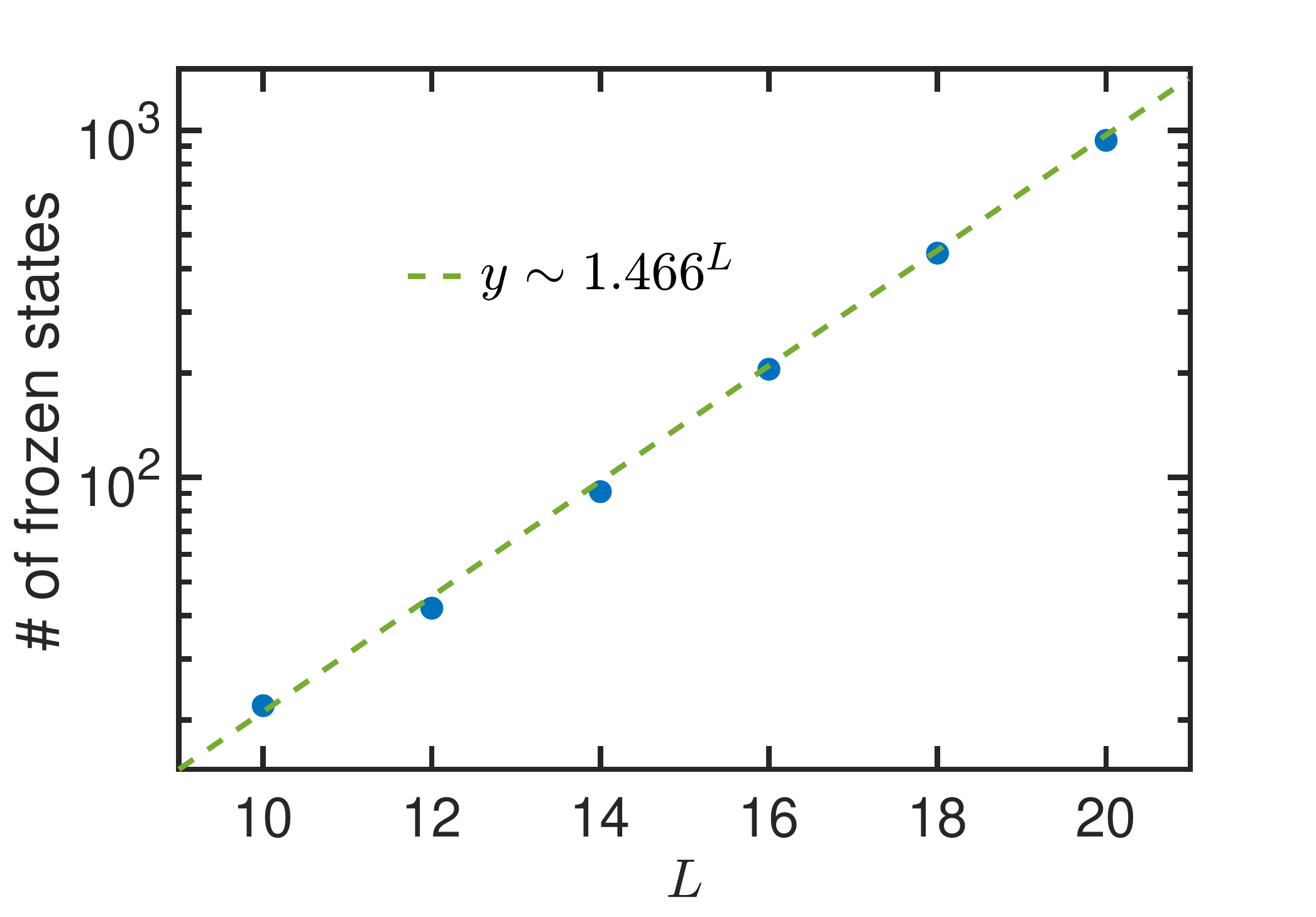}
\caption{Scaling of the total number of frozen states in $H_{\rm eff}^{(4)}$ as a function of the system size. The result agrees with the scaling form $y\sim 1.466^L$.}
\label{fig:frozen_fourth} 
\end{figure}

One can carry out a similar inductive counting procedure as outlined in the previous section of this Supplemental Material. However, if one is only interested in the asymptotic behavior in the limit of large system size $L$, one can show that the frozen space subspace satisfies a generalized Fibonacci constraint which we call the ``Narayana constraint". Since under $H_{\rm eff}^{(4)}$, domain walls separated by distance two are no longer frozen, the new constraint now becomes: there cannot be two next-nearest-neighbor domain walls in the frozen subspace. Let us denote a domain wall by $|\bullet\rangle$, and the absence of a domain wall by $|\circ\rangle$. If a frozen configuration of size $L$ has its boundary in state $|\cdots \circ\rangle$, it could have been obtained by appending $\circ$ to any frozen state of size $L-1$. However, if its boundary is in state $|\cdots \bullet \rangle$, it can only be obtained by appending $\circ \circ \bullet$ to a frozen state of size $L-3$. Therefore the Hilbert space dimension of the frozen subspace grows according to $d_L = d_{L-1} + d_{L-3}$, which is known as the Narayana sequence. The asymptotic behavior of this sequence can be obtained from the characteristic polynomial, from which we obtain $d_L \sim 1.466^L$. Numerical verfication of this scaling is shown in Fig.~\ref{fig:frozen_fourth}, where we find good agreement.

\end{document}